\documentclass[12pt,a4paper]{article}
\usepackage{graphicx,color}
\usepackage{dcolumn,amsmath,amssymb}
\usepackage{wrapfig}

\usepackage{url}

  \newtheorem{theorema}{Theorem}
\newtheorem{problema}{Problem}
\begin{document}
\newcounter{nmiton}

\title{New Error Tolerant Method to Search Long Repeats in Symbol Sequences}

\author{\textsl{Sergey P.\,Tsarev$^1$, \quad Michael G.\,Sadovsky$^2$}\\
${}^1$Siberian Federal University, Institute of space and information technologies;\\ 660026 Russia, Krasnoyarsk, Kirenskogo str., 26. \textit{sptsarev@mail.ru};\\
${}^2$Institute of computational modelling of SD of RAS;\\ 660036 Russia, Krasnoyarsk, Akademgorodok. \textit{msad@icm.krasn.ru}}
\maketitle

\begin{abstract}
A new method to identify all sufficiently long repeating substrings in one or several symbol sequences is proposed. The method is based on a specific gauge applied to symbol sequences that guarantees identification of the repeating substrings. It allows the matching of substrings to contain a given level of errors. The gauge is based on the development of a heavily sparse dictionary of repeats, thus drastically accelerating the search procedure. Some genomic applications illustrate the method.

This paper is the extended and detailed version of the presentation at the $3^{\textrm{rd}}$ International Conference on Algorithms for Computational Biology to be held at Trujillo, Spain, June 21--22, 2016.
\end{abstract}\bigskip

\noindent\textbf{Keywords}: genome, fast search, Vernier pattern, substring search\medskip

\noindent\textbf{PACS}: {87.10.Vg, 87.14.G-, 87.15.Qt}\medskip

\noindent\textbf{MSC}: 11K31, 60-08, 62-07, 68Q25, 68U15, 68W99, 92-08, 92D25\medskip

\setcounter{footnote}{0}

\section{Introduction}
The classic problem of the search of sufficiently long common substring in two symbol sequences has a long story \cite{1,3,4,5,7,8}. In spite of a number of deep and valuable results \cite{Red,Erciyes,RepFindAlg,prap,repexplorer} obtained in the algorithm implementation for the problem, it is computationally challenging and an extremely active research field.

In brief, the problem we address here is the following. Let one (or more) sequences $\mathfrak{T}_1$, $\mathfrak{T}_2$, \ldots , $\mathfrak{T}_k$ from some finite alphabet are given; further we shall concentrate on the four-letter alphabet $\aleph=\{\mathsf{A}, \mathsf{C}, \mathsf{G}, \mathsf{T}\}$ only, since we illustrate the results with genetic data. So, the problem is to find all sufficiently long substrings $\{\mathfrak{s}_i\}$ that occur at least twice in one or several $\mathfrak{T}_i$.
The problem could be understood in two different ways: the former is a search for the \textbf{exactly} matching substrings, and the latter is a search for two substrings bearing some tolerable mismatches; obviously, the first problem is a special case of the second one. Section~\ref{grubo} describes a primitive search algorithm meeting the exact match constraint; Section~\ref{sec-vernier-all} presents our main idea of much faster search method which additionally allows an expansion for approximate matching case. In Section~\ref{sec-experiments} we give a brief survey of experimental verifications of our method. In Section~\ref{sec-disc} an important combinatorial problem related to the proposed method is discussed. The new method has the following advantages:
\begin{list}{--}{\leftmargin=6mm \labelwidth=5mm \topsep=0mm \labelsep=2mm \itemsep=1pt \parsep=0mm \itemindent=10pt}
\item it is much more economic in comparison to exhaustive search for all repeating substrings of an arbitrary length. We search for all repeats longer than a given integer $N$ (provided by researcher). Greater $N$ accelerates our algorithm;
\item it finds simultaneously all repeats in a given DNA sequence (or in any other string of symbols in any finite alphabet) or common substrings in two or more symbol sequences;
\item it permits an error tolerance: a portion of mismatches in compared substrings is allowed. Although the current implementation does not guarantee all repeats (or common substrings) of the length $N$ or greater with given tolerance level of mismatches identification, test runs have shown that the probability of missing of inexact repeats with the given tolerance is small. Still the current implementation guarantees all exact repeats to be found, cf.~the discussion in Section~\ref{metodsam};
\item currently implemented version of the method under consideration allows neither insertions, nor deletions, in the sequences to be compared. Later, we present a new version of the method free from this constraint.
\end{list}
There is a number of various algorithms to resolve the problem, and the number of software implementations falls beyond imagination. Nonetheless, some related results and reference details could be found in \cite{Red,Erciyes,RepFindAlg,prap,repexplorer}. The papers \cite{Red,RepFindAlg} present few versions of well-known algorithms to support the search, identification and sorting of the strings found in a text (see also \cite{2greka}). Rather good and comprehensive survey of the methods and their software implementations is provided in \cite{Erciyes}. Some minor while informative progress in repeats finding and analysis is reported in \cite{prap,repexplorer}.

The search for the exactly matching strings in two (or several) sequences is of great importance; meanwhile, a search of the similarities allowing some (minor) mismatches is of the greatest application interest. Such interest comes both from technical reasons (reading and sequencing mistakes), and from essential ones (detection of the evolutionary changes manifested in symbol mutations in various genetic entities). Here classical paper by D.\,Sankoff \cite{sank72} is still very important and relevant.

What concerns the specific problem of the search for the longest common substring, still there is a lot of papers. The problem and some approaches are discussed in \cite{2greka,Namikietal,taba}; technically advanced tool relevant to the discussed issue is reported in \cite{Nguyenetal}, while some basic algorithmic details and ideas are present in \cite{BMCAlgoritm,bioinf08}.

Another approach dealing with implementation of parallelism into the methodology of a similarity search, as well as the software design is discussed in \cite{bioinf14}. The relevant problem of the assessment of the quality of the identification of those proximal similarities is discussed in \cite{BBRC10}. Finally, some mathematically oriented papers \cite{IEEE2014,kangning} should be mentioned.

Everywhere further we shall concentrate on DNA sequences analysis, while other applications are possible (\cite{Znamenskij2014}). Keeping in mind the double-strand DNA structure, we shall not hereafter discuss repeats search in two strands; it brings nothing from the point of view of the method idea and practically is just a matter of a sequence pretreatment.

\section{Long Repeats by Brute Force}\label{grubo}
Here we sketch a well-known primitive but exhaustive algorithm to search repeating substrings in symbol strings. This algorithm is still of practical use for analysis of DNA sequences as long as $10^7$ or so,  and can be used later to check more advanced algorithms described below.

Theoretically the problem of searching for a repeating substring may be reduced to a construction of a frequency dictionary for the given symbol sequence~$\mathfrak{T}$; the former is a list of all the substrings (of the given length~$m$, also called ``thickness'' of dictionary) occurred within the sequence~$\mathfrak{T}$ so that each entry in the dictionary is associated with the frequency of the relevant string in $\mathfrak{T}$. Frequency dictionary~$W_m$ (of the thickness~$m$) is the key object of the studies in a variety of fields ranging from pure mathematics to bioinformatics and linguistics.
A dictionary~$W_m$  (of the thickness~$m$) could be defined in a variety of ways; cf.~ for example~\cite{sadov3} (where it is called \textsl{finite dictionary}). Note that the definition of the frequency dictionary used here differs slightly from the common one, see~\cite{sadov3},

The simplest way to develop $W_m$ is as follows. Let us fix a window of the length~$m$ that identifies a substring in a sequence~$\mathfrak{T}$, and a step~$t$ for the window shift alongside the sequence. Thus, the frequency dictionary $W_{m,t}$ is the set of all the strings of the length~$m$ identified by the window of that length moving alongside the sequence with the step~$t$. Each element of the dictionary is assigned with its frequency (the number of copies of this element met in the dictionary building process) and (for our purposes) the list of all positions where the given element of the dictionary has been met. Practically, $t=1$ almost always; it means that each symbol in a sequence gives a start for substring (of the length~$m$) in $W_m$.

Having $W_{m,1}$, one easily can find all the repeats of the length $N\geqslant m$ in $\mathfrak{T}$, selecting all elements $s$ of $W_{m,1}$ that are met more than in one copy and (using the list of the positions of $s$) clustering all other repeating elements of $W_{m,1}$ with consecutive position tags.

The question whether two (or several) sequences~$\mathfrak{T}_1$ and~$\mathfrak{T}_2$ have a common substring~$s$ of the length~$N$ could be addressed through a comparison of the frequency dictionaries of those sequences. Obviously, one must develop a series of dictionaries $W^{j}_{m,1}$ for some convenient $m$, $1 \leqslant m \leqslant N$; index~$j$ corresponds to the sequence~$\mathfrak{T}_j$. Then, one should compare the dictionaries, and the comparison always yields the common substrings of length $N \geqslant m$ to be found in~$\mathfrak{T}_j$. This procedure could be completed in a finite (maybe rather long) time, and always brings a result. Yet, it is rather laborious and require proper hardware to carry it out.

In our test runs we have built up dictionaries of thickness $m \leqslant 10000$ for a single DNA sequence of length $44\cdot 10^6$ base pairs (\emph{Bos taurus} chromosome 25). The dictionaries of the given thickness were built in three stages:
\begin{list}{\arabic{nmiton})}{\usecounter{nmiton}\leftmargin=6mm \labelwidth=5mm \topsep=0mm \labelsep=2mm \itemsep=1pt \parsep=0mm \itemindent=10pt}
\item first, we identified substrings of the given length $m$ with step $t=1$ and develop an intermediate ``predictionary'' text file \verb"F.predic" where each substring occupies a separate line and is accompanied by the position tag;
\item second, \verb"F.predic" is sorted lexicographically using the standard system command \verb"sort";
\item third, the identical substrings in the sorted file are eliminated so that the resulting line bears the substring, accumulated number of its copies and the list of position tags gathered from the eliminated substrings in the sorted \verb"F.predic".
\end{list}

\setlength{\intextsep}{1pt}
\begin{wraptable}{l}{8cm}
\caption{\label{sfu_cluster} Runtime of the tests for brute-force dictionary development; $t_{\textrm{s}}$ is the sorting time, $m$ is a substring length.}
\begin{tabular}{|r|c|c|}\hline
\multicolumn{1}{|c|}{$\qquad m\qquad $} & \verb"F.predic" size & $t_{\textrm{s}}$ \\ \hline \hline
200 & 8.3 Gbytes & 10 min \\ \hline
500  & 20 Gbytes & 11 min \\\hline
1000 & 40 Gbytes & 17 min \\\hline
10000 & 400 Gbytes & 1 h. 12 min \\\hline
\end{tabular}
\end{wraptable}
Stage 2 (lexicographical sorting) is the most time consuming. It could be executed in reasonable time on a mainframe with 30~Gb of RAM under OS Linux (\url{http://cluster.sfu-kras.ru/page/supercomputer/}). Table~\ref{sfu_cluster} shows the run time of this step for several $m$ values.

Hence, the steps 1--3 yield the following results, in terms of the frequency dictionary structure, see Table~\ref{metka}, where the observed number $N_k$ of (different) strings (of the length $m$) met in $k$ copies are given.
\begin{table}[h]
\caption{\label{metka}Abundance of frequency dictionary, 
$m$ is the substring length.}
\begin{center}
\begin{tabular}{|p{1.3cm}|c|c|c|c|c|c|c|c|c|}\hline
\multicolumn{10}{|c|}{$m=200$}\\\hline
$k$ & 2 & 3& 4 & 5 & 6 & 7 & 8 & 9 & $\geqslant 10$\\\hline
$N_k$ & 312338 & 3600 & 756 & 203 & 72 & 2 & 0 & 0 & 80\\\hline
\multicolumn{10}{|c|}{$m=500$}\\\hline
$k$ & 2 & 3& 4 & 5 & 6 & 7 & 8 & 9 & $\geqslant 10$\\\hline
$N_k$ & 252338 & 126 & 14 & 33 & 0 &0 & 0 & 0 & 33\\\hline
\end{tabular}
\end{center}
\end{table}

For $m=1000$ $N_2=193227$ strings have been met in two copies, and none has been found in three or more copies. Similarly, for $m=10000$ these figures were $N_2=18865$ and $N_{>2}=0$, respectively. In fact all repeats of substrings of length 10000 in this DNA sequence were clustered on step 3 into 3 exactly matching substrings of lengths  (approximately) 11000, 15000 and 21700.

\section{Vernier Gauge Algorithm}\label{sec-vernier-all}
Here we introduce a new much faster method to search for the longest common substring in two sequences.
\begin{problema}Assuming an occurrence of a repeating substring of the length $\geqslant N$ to be found in a symbol sequence (alternatively, a common substring in  several symbol sequences), can we do it \textbf{much faster} than the brute force method (or its versions) does?
\end{problema}

The way to resolve the problem positively is shown below. The key idea to do the search faster consists in a change of a complete frequency dictionary~$W_{m,1}$ (where each symbol in the sequence $\mathfrak{T}$ gives a start to a string of the length~$m$) for a sparse frequency dictionary $W_{m,t}$ with \emph{variable step $t$}; the dictionary $W_{m,t}$ has significantly less number of entries. An idea standing behind the proposed method is strongly connected to a well-known Vernier scale \cite{Vernier-wiki} used to measure length with enhanced precision in comparison to the standard scale.

\subsection{Simple Example}\label{simprimer}
In this subsection we develop the idea of the \emph{simplest} Vernier gauge to search a common string of length $N$ or more in two symbol sequences. Simply speaking, we should cover the first sequence with  \textsl{tags} of some small length $m$ with some step $k$; the second sequence must be covered with the \textsl{tags} of the same length, but here  the step between (the starting letters of) two neighboring tags is equal to $k-1$, but not $k$. If a tag is found in both sequences, it must be examined for \emph{expansion} (see below). Let us give a closer look at this process.

Suppose two sequences~$\mathfrak{T}_1$ and~$\mathfrak{T}_2$ have a common string~$\mathfrak{s}$ of the length~$N$; yet, we have no idea about the locations of that common string in the sequences. Let us build two frequency dictionaries: $W^1_{m, k}$ and $W^2_{m, k-1}$ with $k \leqslant \sqrt{N}$ for $\mathfrak{T}_1$ and~$\mathfrak{T}_2$ respectively. These two frequency dictionaries are of the some thickness $m$ (that is the length of strings enlisted at a dictionary). Both dictionaries take start in the development from the very beginning of each sequence.
\begin{figure}
\includegraphics[width=16.5cm]{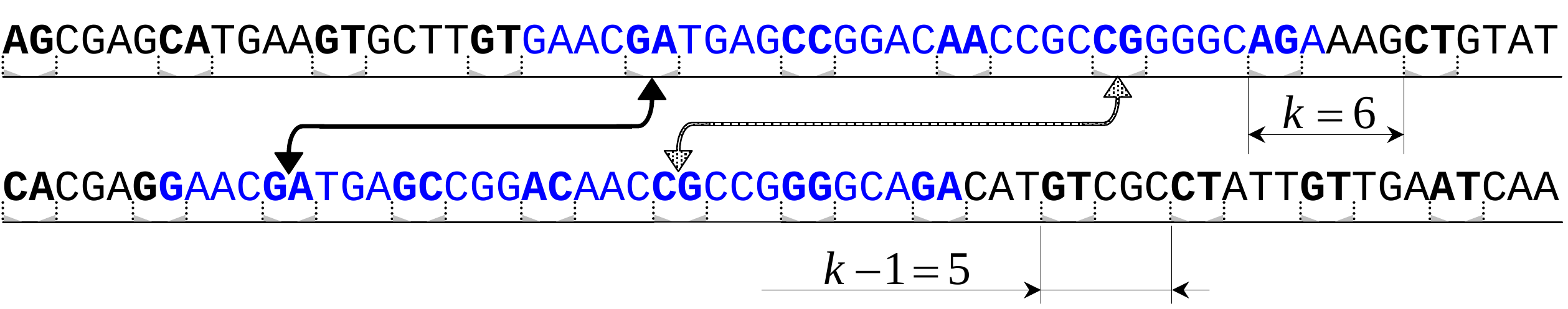}
\caption{\label{fig1} Illustration of the Vernier approach to find out a sufficiently long common substring in two sequences.}
\end{figure}

Fig.~\ref{fig1} illustrates this idea for $k=6$, $m=2$. The common string of length $N=31$ is indicated in \textcolor{blue}{blue}, in the figure. The Vernier gauge (see Section~\ref{gd} for details) identifies the common short sub-substring $\mathsf{GA}$ 
in both sequences; they are indicated with the curve arrow in black. Incidentally we can see one more occurrence of a common substring $\mathsf{CG}$ of length~2 in our target common strings; this is not a typical case, but we shall keep in mind such a possibility, as well.

A closer inspection gives three more common substring $\mathsf{CA}$, $\mathsf{CT}$ and $\mathsf{GT}$ of the length~2 in the dictionaries $W^1_{2, 6}$ and $W^2_{2, 5}$. These are not \emph{expandable} to a common target string of the length $N=5\times 6 + 2 -1=31$ in $\mathfrak{T}_1$ and~$\mathfrak{T}_2$ (more on \emph{expansion} of the common entries in the dictionaries $W^1_{m, k}$ and $W^2_{m, k-1}$ see below and in subsection~\ref{metodsam}). This abundance of common entries in $W^1_{2, 6}$ and $W^2_{2, 5}$ results from the small capacity of the nucleotide alphabet $\aleph = \{\mathsf{A,C,G,T}\}$ and a smallness of the length $m=2$ chosen for tags, at this simple example. The choice of $k$ guarantees there always be at least one common string of the length~$m$ to be found in these two dictionaries, if a common substring of length $N\geqslant k(k-1)+m-1$ exists (cf.~Theorem~\ref{teorema1}).

It should be stressed that parameters $k$ and $m$ are almost independent (cf.~Theorem~\ref{teorema1} for a precise statement). The choice of these parameters is determined by the (expected) length $N$ of a common substring, while $m$ is to be chosen almost arbitrary: it would be nice, if the frequency dictionary of the tags is almost degenerated (i.\,e. the greatest majority of the tags should exist in few copies).

The idea to search the common string using relatively short tags and rarefied dictionaries (we call it the  \emph{double Vernier gauge} on $\mathfrak{T}_i$) is based on the following simple theorem.
\begin{theorema}\label{teorema1}
If there is a common string of the length~$N$ or more in~$\mathfrak{T}_1$ and~$\mathfrak{T}_2$ then a common entry (substring of length~$m$) can be found in dictionaries $W^1_{m, k}$ and $W^2_{m, k-1}$ developed for sequences~$\mathfrak{T}_1$ and~$\mathfrak{T}_2$, respectively, provided that $N \geqslant k(k-1) + m -1$.
\end{theorema}

\textbf{Proof}. Let $s_1$, $s_2$ be the common (exactly matching) substrings of the length $N \geqslant k(k-1) + m -1$ starting at positions $u$ and $v$ in $\mathfrak{T}_1$ and $\mathfrak{T}_2$, respectively. First, we cut off (virtually, for simplicity of the proof) their last $m-1$ symbols and look on at the starting positions of the tags, only, at the dictionaries $W^i_{m, k}$ in $s_i$. Let $0 \leqslant \alpha < k$ and $0 \leqslant \beta < k-1$ be the starting positions of the dictionary entries of $s_1$  and $s_2$ \emph{with respect to their starting symbols}, accordingly. The other starting positions of the dictionary tags inside $s$ is $\alpha + x\cdot k$ with $ x \in \{0,1, \ldots , k-2 \}$, for starting point $u$ and $\beta + y\cdot k$ with $ y \in \{0,1, \ldots , k-1 \}$, for starting point $v$. Find then such integers $x$, $y$ that $\alpha + x\cdot k = \beta + y\cdot (k-1)$, i.\,e.\ $x\cdot k -  y\cdot (k-1) = \beta -\alpha $ with the constraints on $x$, $y$ given above. Since $\mathsf{GCD}(k,k-1)=1$, the standard extended Euclid algorithm guarantees the existence of the integers $x$, $y$ meeting the constraints given above.

Here $\mathsf{GCD}(c,d)$ is the greatest common divisor of numbers $c$ and $d$, respectively. In fact, if $\gamma = \beta -\alpha \geqslant 0$, then $x=y=\gamma$ is the solution. For $\gamma = \beta -\alpha < 0$, $x=\gamma+k-1$, $y=\gamma+k$ should be taken.\hfill $\square$

In order to find duplicate tags in the dictionaries one may apply different standard techniques; in our current simplest implementation standard lexicographic sorting and merging of $W^1_{m, k}$ and $W^2_{m, k-1}$ are used.
%
As soon as all common tags in the dictionaries $W^1_{m, k}$ and $W^2_{m, k-1}$ are found, the next steps of our algorithm follow:
\begin{list}{--}{\leftmargin=6mm \labelwidth=5mm \topsep=0mm \labelsep=2mm \itemsep=1pt \parsep=0mm \itemindent=10pt}
\item \emph{expand} the detected common tags starting from their positions in $\mathfrak{T}_1$ and $\mathfrak{T}_2$. Namely, compare consecutively the sequences symbol by symbol to the right of the tags in $\mathfrak{T}_1$ and~$\mathfrak{T}_2$ as long, as they match, stopping when a non-matching symbol is met. Then compare consecutively the sequences symbol by symbol to the left of the tags, in the same manner, as far as they match;
\item if the length of the string obtained through the \emph{tag expansion} is at least $N$, add it to the list of successful expansions for further output, upon the identification of all tag couples in $W^1_{m, k}$ and $W^2_{m, k-1}$ that could be \emph{expanded}.
\end{list}
The simplest version of our algorithm discussed here stipulates the search for \emph{exactly matching} strings, only. Also we do not take into account the possibility to meet some other symbols than the standard nucleotides $\mathsf{A}$, $\mathsf{C}$, $\mathsf{G}$, $\mathsf{T}$. If one expects that some other symbols (like $\mathsf{N}$, $\mathsf{W}$ etc.\footnote{\url{http://www.chem.qmul.ac.uk/iubmb/misc/naseq.html}}) may occur in analyzed DNA sequences, then an expansion strategy from those discussed in Section~\ref{metodsam} should be applied. The choice of the parameter $m$ is considered in the next subsection.

\subsection{Tag Length Choice to Enforce Vernier Gauge Algorithm}\label{sec-tag}
The substrings of the length $m$ chosen to build up the dictionaries $W^1_{m, k}$ and $W^2_{m, k-1}$ in the previous subsection are called \textbf{tags}. Its length~$m$ is very important parameter affecting speed and overall efficiency of the algorithm. A smart choice of this parameter may dramatically reduce the processing time and, what is even more important, is crucial in the process of subsequent \emph{expansion} of the tags common for both dictionaries $W^1_{m, k}$ and $W^2_{m, k-1}$ toward the full common strings of the length $\geqslant N$ ($N\geqslant k(k-1)$) \textsl{with the given portion of errors} at the last stage of the algorithm execution (Section~\ref{gd}).

In fact, the capacity of DNA alphabet dictates the choice of sufficiently large~$m$ to minimize the number of sporadic coincidences of tags in the dictionaries. The experiments presented in Section~\ref{grubo} show that $m=1000$ practically guarantees that such long tags occurring in both dictionaries $W^1_{m, k}$, $W^2_{m, k-1}$ will expand to a longer common string. But, further results (see Section~\ref{sec-experiments} below) show that even much shorter $m=30$ is good enough, when the steps $k$, $k-1$ are greater than~30 (so the target length of common substrings sought by the algorithm is at least~1000). It should be stressed that overlapping of the tags in $\mathfrak{T}_1$ and $\mathfrak{T}_2$ (if $m>k-1$) is not a problem, for our approach; overlapping itself does not affect the algorithm, cf.~\ref{teorema1}.

As soon as the main parameter $N$ of the algorithm (the minimal length of common substrings in $\mathfrak{T}_1$ and $\mathfrak{T}_2$ we are looking for) is fixed, the choice of $m$ also affects the derivative parameters $k$ and $k-1$. Rigorously speaking, to guarantee that all common strings of the length $\geqslant N$ in $\mathfrak{T}_1$ and $\mathfrak{T}_2$ are found, one must choose $k$ and $m$ so that $N \geqslant k(k-1)+m-1$.

\subsection{General Description of the Problem}\label{gd}
The double Vernier gauge described in Section~\ref{simprimer} stands behind the more general search pattern presented below in Sections~\ref{krug}, \ref{sec-min-opt}. Here we discuss the simplest modifications of the double Vernier gauge necessary to find repeats in one or several DNA sequences, only.

The general problem solved with our algorithms sounds as following:
\begin{problema}
Given parameters $N$ (an integer) and $\varepsilon$ (a positive real number), find substrings of the length at least $N$ in one or several sequences $\mathfrak{T}_i$, $i = 1, 2, \ldots , \,t$ that occur repeatedly (exact matching requires $\varepsilon=0$) or couples of strings in $\mathfrak{T}_i$ differing at most at $q$ positions, $q=\left[\varepsilon\cdot \mathrm{length}(s)\right]$.
\end{problema}
Here $[x]$ denotes the integer part of $x$.

\subsection{How the Method Works}\label{metodsam}
This is how the method works.
\begin{list}{\textsl{Step} \arabic{nmiton}. }{\usecounter{nmiton}\leftmargin=6mm \labelwidth=5mm \topsep=0mm \labelsep=2mm \itemsep=1pt \parsep=0mm \itemindent=10pt}
 \item Given the target length $N$, choose  proper $k$ and $m$ so that  $N \geqslant k(k-1)+m-1$.
 \item If two DNA sequences are analyzed, develop the dictionaries $W^1_{m, k}$ and $W^2_{m, k-1}$. Otherwise (for one or more than two DNA sequences) develop for each DNA sequence a dictionary with variable step: take tags of length $m$ starting at positions $1$, $k$, $k+1$, $2(k-1) + 1$, $2k +1$, \ldots that is at the union of the subsets $\{p\cdot (k-1) +1,  p = 0, 1, 2, \ldots\}$ and $\{q\cdot k + 1,  q = 0, 1, 2, \ldots\}$. Add positions of the selected tags into the dictionaries.
 \item  Check whether there may be common entry tags in the dictionaries. If we want to find repeats in one DNA sequence (or all possible repeats in several DNA sequences $\mathfrak{T}_i$, possibly the same ones), find repeated tags in one dictionary or in the dictionary merged from all dictionaries built for all $\mathfrak{T}_i$. Strategies for finding common tag entries are discussed below.
\item \emph{Expand} the detected repeating tags (using their positions stored in the dictionaries) as described in Section~\ref{simprimer} if the exact matching is required. When a positive tolerance $\varepsilon > 0$ is allowed, use one of the expansion strategies discussed below.
\item List all expanded tags with their positions in $\mathfrak{T}_i$. If some of the expanded substrings are shorter than $N$ one may keep or reject them (there is no guarantee that all matching substrings of length $< N$ will be found!)
\end{list}

\subsection{More technical details}
\emph{Finding common tags on Step 3}. Currently, we use the standard lexicographic sorting of the tags (using the standard system command \texttt{sort}) followed with merging the sorted dictionaries. Each entry of the dictionary consists of a tag and the relevant positions of that latter, so sorting brings the identical tags into a series of consecutive lines. Using any text processing utility (for example the standard \texttt{gawk}), we gather such consecutive lines with identical tags into a single one thus building a list of positions of a given tag, in the relevant $\mathfrak{T}_i$. This procedure is rather fast for the examples described in Section~\ref{sec-experiments}; note that for error tolerance $\varepsilon>0$ (so inexact matches are allowed) lexicographic sorting does not guarantee that all the tags matching inexactly with the given tolerance level  $\varepsilon>0$ would be found. In this case more advanced string matching algorithms should be applied. In fact, if $m$ is significantly less than $N$ (in our experiments $\varepsilon=1/50$, $m=30$ or so and $N \geqslant 1000$), the probability to shoot a tag into inexactly matching site 
on Step~2 is rather small. So even straightforward sorting makes a reasonable choice; one may repeat the Steps~2,~3 with shifted positions of the tags (several shifts of order $m$ are recommended). It will increase the probability to detect all inexactly matched strings of target length $N$ or longer.

\emph{Expansion strategies on Step 4}. If $\varepsilon=0$, simple expansion described in  Section~\ref{simprimer} should be applied, that is a consecutive comparison of the symbols on the right and on the left from the identical tags in $\mathfrak{T}_i$, as far as they match. The expansion stops as one meets a non-matching symbol. For $\varepsilon>0$, the expansion goes on even if the symbols located near the tags do not match;
if such non-matching is found, add~1 to the counter \texttt{miss\_count} of mismatches and stop the expansion process as soon, as $\texttt{miss\_count}/\mathrm{length}(s) > \varepsilon$ (here $s$ is the string obtained in the expansion process). A good idea here is to repeat the expansion several times to the left and to the right from the matching tags, since $\mathrm{length}(s)$ grows up in different manner, for various directions of expansion.

\emph{Processing of symbols $\mathsf{N}$, $\mathsf{W}$ etc. falling out the original alphabet $\aleph$.} Up-to-date DNA databases contain a lot of sequences with non-exact recognition of nucleotides. Such misrecognitions are denoted by letters falling outside the standard DNA alphabet $\aleph=\{\mathsf{A}, \mathsf{C}, \mathsf{G}, \mathsf{T}\}$. Several strategies may be applied here depending on the problem to be solved by a researcher:
\begin{list}{--}{\leftmargin=6mm \labelwidth=5mm \topsep=0mm \labelsep=2mm \itemsep=1pt \parsep=0mm \itemindent=10pt}
  \item consider symbols $\mathsf{N}$, $\mathsf{W}$ etc. as \emph{errors} adding 1 to \texttt{miss\_count};
  \item consider them as \emph{possible matches} (not recognized by the DNA sequencer) and keep expansion without  adding 1 to \texttt{miss\_count}. It may result in very long expansions consisted mostly of $\mathsf{N}$'s; hence, one has to examine the results of an expansion, or apply other criteria to stop an expansion;
  \item cut the DNA sequences into smaller pieces free from those extra symbols, and run the algorithm on the obtained pieces. In practice, such strategy results in generation of thousands of separate files, in majority of cases. Our experiments show that it is not a problem for our implementation and possibly for any other reasonable choice of text processing routines on Steps~1--4.
\end{list}

\section{Preliminary Experimental Results}\label{sec-experiments}
We checked the developed algorithm over the following genetic data (all sequences were retrieved from EMBL--bank\footnote{\url{http://www.ebi.ac.uk/genomes}}):
\begin{list}{\arabic{nmiton})}{\usecounter{nmiton}\leftmargin=6mm \labelwidth=5mm \topsep=0mm \labelsep=2mm \itemsep=1pt \parsep=0mm \itemindent=-1pt}
\item \emph{Human chromosome 14} (since it contains $\mathsf{A}, \mathsf{C}, \mathsf{G}, \mathsf{T}$ symbols only);
\item 4 sets of drosophila genomes:
\begin{itemize}
  \item \emph{Drosophila melanogaster},
  \item \emph{Drosophila simulans},
  \item \emph{Drosophila simulans} strain white501,
  \item \emph{Drosophila yakuba} strain Tai18E2;
\end{itemize}
\item \emph{Bos taurus} complete genome.
\end{list}

\subsection{Human chromosome 14}\label{sec-homo}
An execution of the algorithm described in Section~\ref{metodsam} with the parameters $m=50$, $k=31$ (so we find all repeats of length $N\geqslant k(k-1)+m-1=979$ brings 19946 repeated tags, totally, at the Step~3. Among them, 12154 tags occur twice, 3670 tags occur thrice, \ldots, 205
tags occur 10 times or more, and the maximal frequency of 25 was found, for the tag {\tt{ctttctttctttctttctttctttctttctttctttctttctttctttct}}.

An expansion according to Step~4 with $\varepsilon=0$ (only the exact matching was allowed) yielded two identical strings of length 1019, as well as hundreds of shorter repeats. An expansion according to Step~4 with $\varepsilon=0.02$ yielded a couple of approximately matching strings of the length 11000. This is an approximate length, since the expansion method used here brought a few dozens of mismatches at the both ends of them, as well as few hundreds of inexact repeats of the lengths 1000 and more. The couple with length $>11000$ is in fact a long almost periodic subsequence with period~102: the first string at the couple starts from position 85\,597\,640 and the second one was shifted to the end of the chromosome by~102 positions. A comparison of these approximately matching strings reveals a few \emph{exactly matching substrings} of the following lengths:
38,
101,
305,
203,
468,
101,
652,
101,
298,
55,
38,
62,
242,
94,
196,
101,
101,
203,
108,
305,
203,
305,
101,
101,
94,
101,
196,
94,
101,
62,
38,
62,
344,
62,
38,
62,
140,
62,
38,
62,
242,
101,
196,
94,
101,
24,
123,
52,
24,
21,
77,
287,
94,
196,
203,
62,
38,
62,
147,
101,
157,
203,
305,
101,
45,
101,
203,
94,
196,
713,
101,
62,
101,
38,
62,
38,
164 (they are given in the order of appearance; the exact matches of length $>10$ only are shown). Many of them are the multiples to the period~102 minus~1. Typically, a single mismatching nucleotide only occurs between the exactly matching subregions. This pattern of inner exactly matching substrings is rather typical over the results of our computational experiments. Hence, the choice of the tolerance level $\varepsilon=1/50$ seems reasonable for the genomes under consideration.

This longest approximate repeat has few other interesting features. For example, the exact repeat of the length $1019=102\cdot 10-1$ actually occurs inside the longest couple of approximate repeats, if the comparison was carried out with the shift $3366=102\cdot 33$. Such multiply repeated 102~symbols long substring occurred at this subregion of the length~019 nucleotides is following: \texttt{tgggggcgggggacagcctggcagccccgtggcaccctcaggagcaa caacctagcatctcaggagagagaggccacaccactgtccgcgtagtcgcccagc}. 

\subsection{Four \textit{Drosophila} genomes}\label{sec-dros1}
The collection of 24 chromosomes of all 4 species consists of $478\cdot 10^6$~bp; it comprises several millions of unrecognized nucleotides (marked \verb"n", mostly met in the last 3 species genomes). An execution of the algorithm of Section~\ref{metodsam} with the parameters $m=50$, $k=63$ (so we expected all repeats of the length $N\geqslant 3995$) over the complete set of all chromosomes revealed in total 180980 repeated tags of the length~$m$ on Step~3, with the maximal frequency equal to 34 for the tag \texttt{ataataataataataataataataataataataataataataataataat}.

We tried another 1parameters $m=50$, $k=200$ (hence expecting to find out all repeats of the length $N\geqslant 40000$) and obtained 20442 repeated tags (with maximal frequency equal to 12). On Step~4, we treated \verb"n" symbols as errors. For $\varepsilon=0$, 7 repeats of the length $\geqslant 100000$ (within a given organism genome, and between them) were found, one of them is 177722~bp long. This exact match also includes a considerable portion of \verb"n" symbols. Hundreds of other exact repeats of the length $\leqslant 10000$ were also revealed.

For $\varepsilon=1/50$, the exactly matching couples found in the previous experiment showed further expansion; for example, the exact repeat of the length 177722 generated a longer approximate repeat. When treated separately, four \textit{Drosophila} species exhibit considerable variation in repeat lengths, and the overall numbers of long repeats (for $\varepsilon=1/50$) are:
\begin{list}{--}{\leftmargin=6mm \labelwidth=5mm \topsep=0mm \labelsep=2mm \itemsep=1pt \parsep=0mm \itemindent=-1pt}
  \item \emph{Drosophila melanogaster} genome has 9 repeats of length 10000 and more, the longest one is 30893 nucleotides long.
  \item \emph{Drosophila yakuba} strain Tai18E2, \emph{Drosophila simulans} and \emph{Drosophila simulans} strain white501 genomes do not have such long exact repeats (while they have dozens of exact repeats of the length 1000 and more with the maximal length of 3024 nucleotides). On the other hand, these exact repeats yielded an expansion into approximate repeats of the length up to 6000, when processed with  $\varepsilon=1/50$.
\end{list}

\subsection{\emph{Bos taurus} complete genome}\label{sec-dros2}
The overall size of 29 processed DNA sequences of the complete genome is more than $2.4\cdot 10^9$ symbols. Since the files contain large unrecognized nucleotide substrings, a number of different strategies described in Section~\ref{metodsam} were implemented. Considering the \verb"n" symbols (no other unrecognized symbols were encountered) as non-erroneous ones, one faces few huge repeats of the length up to 300000; they consist of \verb"n" symbols practically completely. When the files were cut into pieces free from \verb"n" symbols (thus yielding 9718 files of 100~Kbytes size and $> 11000$ files of smaller size), then processing with the parameters $m=50$, $k=600$ (and $N>360000$, respectively) brings a number of exact repeats of lengths $\leqslant 89453$.

\section{Discussion}\label{sec-disc}
The experiments described above show that our method is sufficiently fast and yields the results interesting both for exact and approximate sequence analysis.

Still, a number of questions arises concerning feasibility of the method for various biologically meaningful issues; a search for degenerated motifs is among them. First, here we present a theoretical result rather than a ready-to-use software package. Our implementation aims just to check feasibility of the method itself. Evidently, there is no obstacles to combine, in some way, Vernier sparse search and other well-established techniques (suffix trees, etc.).

Second, the current implementation guarantees revealing of all \emph{exact} matches. If the tolerance level $\varepsilon>0$, then the above presented  method remains feasible for inexact repeats search, while some minor changes must be implemented to avoid a failure of the method caused by the coincidence of a tag with admissible mismatches in degenerated motif. Indeed, simple lexicographically arranged sorting of tags (sparse dictionary entries) must be changed for the search of tags that are close with respect to admissible mismatch patterns (Levenshtein distance, edit distance, and other versions of that former taking into account insertions and deletions). Moreover, insertions and deletions admission would result in serious modification of original Vernier gauge.

Finally, a correct comparison of the speed of execution of software implementing Vernier method, and the combinations of that latter with some other approaches should be done explicitly; yet, a number of experiments should be done, as well as the issue of the problem must be clarified.

In addition, some interesting mathematical and algorithmic issues are to be urged to optimize the process. The following subsections address the issues.

\subsection{Circular and Linear Vernier Patterns}\label{krug}
Actually, the procedure described in Section~\ref{sec-vernier-all} implies the following property of the standard double Vernier gauge:
{\bf \begin{quote}
Suppose some positions $i_1$, $i_2$, \ldots , $i_k$ in $N$-element set $\mathfrak{N} = \{1,2, \ldots , N \}$ are marked. Then, if these marks are periodically repeated in a larger set  $\mathfrak{M} = \{1,2, \ldots , M \}$, $M \gg N$, then for any $s <M-N$ one finds at least two marked positions with the exact distance $s$ between them.
\end{quote}}
This property guarantees that for any two identical strings $\mathfrak{s}_1$, $\mathfrak{s}_2$ of the length $N+m-1$ located in a longer symbol sequence $\mathfrak{T}$ of the length $M+m-1$ (the starting positions of $\mathfrak{s}_1$ and $\mathfrak{s}_2$ differ in $s$ symbols), the couple of the marked positions exists in $\mathfrak{T}$ at the same position in $\mathfrak{s}_1$, $\mathfrak{s}_2$ with respect to their starting symbols, so the tags (substring of the length $m$) starting at the selected positions inside  $\mathfrak{T}$ coincide.

A better geometric insight into this \textbf{Vernier pattern} of positions  $i_1$, $i_2$, \ldots , $i_k$ is given by the following construction:
{\bf \begin{quote}
taking a circumference of length $N$ and starting from some point $O$ (corresponding to the element 1 in the set  $\mathfrak{N}$), mark clockwise the points at the distances $i_1-1$, $i_2-1$, \ldots , $i_k-1$ from $O$ on the circumference. Then for any integer length $s \leqslant N/2$ one finds at least two marks spanning the (shortest) arc of length $s$.
\end{quote}}

\begin{wrapfigure}{l}{5cm}\includegraphics[width=4cm]{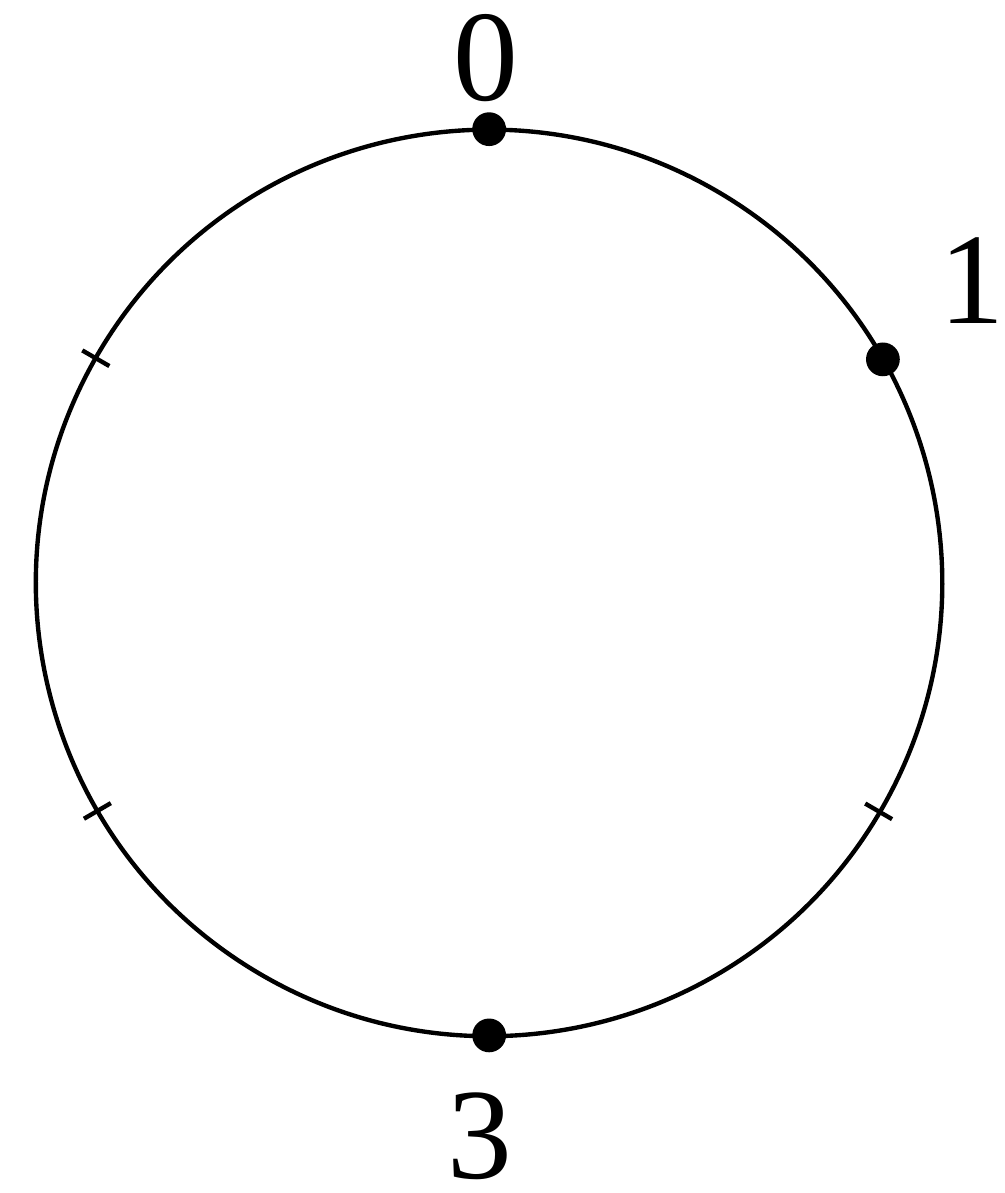}
\caption{\label{fig2}An example of minimalistic circular Vernier patterns, for $k=3$.}
\end{wrapfigure}
The circular picture corresponds to periodic repetition of the marks in the larger $M$-element set $\mathfrak{M}$. If for the given integer $N$  one finds a set of positions  $\mathfrak{V} = \{i_1, i_2, \ldots , i_k\}$ (with integer elements $0<i_p\leqslant N$) meeting the constraints formulated above, then such $\mathfrak{V}$ is called an $(N,k)$-\textbf{circular Vernier pattern}.

Circular Vernier patterns are adapted for a search of repeats of the length $N$  in (much longer) sequences of the length $M >N$.
Another combinatorial problem of a search of all possible pairs of strings $\mathfrak{T}_{i_1}$, $\mathfrak{T}_{i_2}$ from a very large finite collection of strings $\mathfrak{T}_i$ (each of them of approximately the same length $N$) so that $\mathfrak{T}_{i_1}$, $\mathfrak{T}_{i_2}$ can partially overlap the substrings from some larger \emph{merging string}, then a new concept of Vernier pattern should be defined. It is referred to simplified version of the problem of DNA sequence assembly.

Namely, if a subset $\mathfrak{V} = \{i_1, i_2, \ldots , i_k\} \in \mathfrak{N}$ exists so that for any integer length $s \leq N$ one finds at least two elements in $\mathfrak{V}$ with distance $s$ between them, then such $\mathfrak{V}$ is called  $(N,k)$-\textbf{linear Vernier pattern}.

If such linear Vernier pattern exists, then marking the respective positions $i_l$ in all of $\mathfrak{T}_i$ (and ignoring their parts after $N$-th position), one sees that for any two such marked strings aligning substrings of a larger merging string, then at least a pair of marked positions coincides in the merging string. So the same idea to cut tags and find identical tags will work along the lines of the algorithm described in Section~\ref{sec-vernier-all}. Certainly we shall take $m$ small enough and all the lengths of the strings $\mathfrak{T}_i$ shall be at least $N+m$.

\subsection{Minimalistic and Minimal Vernier Patterns}\label{sec-min-opt}
Obviously, the smaller $k$ for given $N$ is taken, the more economic dictionary could be developed using the tags with the starting positions at the elements of an $(N,k)$-circular Vernier pattern $\mathfrak{V}$ periodically repeated in a large DNA sequence $\mathfrak{T}$.

Since the number of different distances between $k$ points can not be greater than $k(k-1)/2$, we have the following lower bound for $k$: $k(k-1)/2 \geqslant N/2$, thus for big $N$, $k \sim \sqrt{N}$. So, for the double Vernier gauge described in Subsection~\ref{simprimer} for a search of repeats in \emph{two} DNA sequences, we have in fact a minimal possible choice of marks (beginning positions of the tags).
\begin{wrapfigure}{r}{5cm}
\includegraphics[width=4cm]{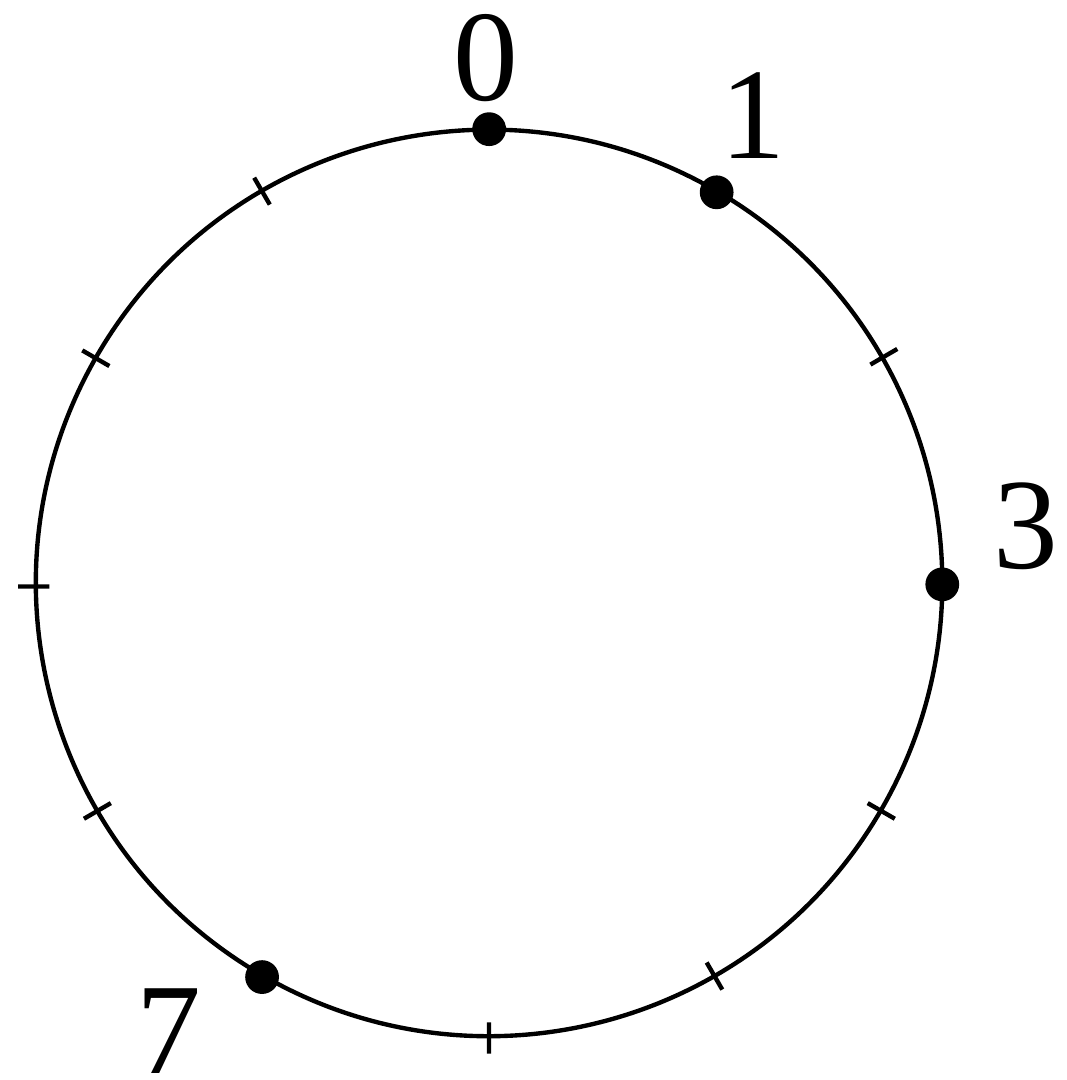}
\caption{\label{fig3}An example of minimalistic circular Vernier patterns, for $k=4$.}
\end{wrapfigure}
For all other cases described as Step~2 of our algorithm in Subsection~\ref{metodsam}, we have approximately twice more marked positions in each of the $\mathfrak{T}_i$. For linear Vernier patterns the situations is slightly different: $k(k-1)/2 \geqslant N_1 = N-m+1$.

So the following mathematical problem makes a good combinatorial challenge.
\begin{problema}
The problem of $\mathfrak{V}$. For any given integer $N$ find circular and linear Vernier patterns with minimal possible $k$.
\end{problema}

\begin{figure}
\hfill\includegraphics[width=3.4cm]{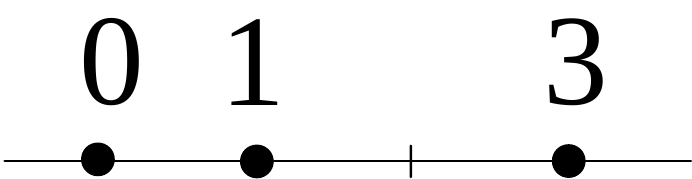}\hfill\includegraphics[width=6cm]{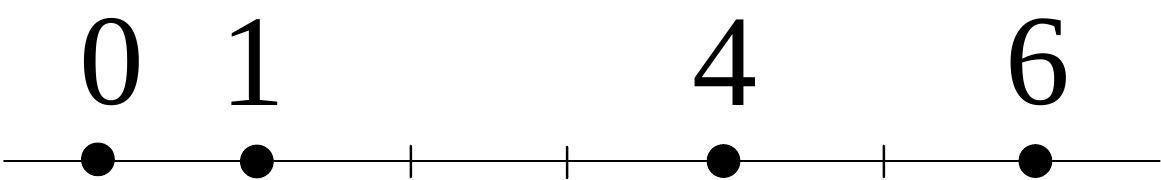}\hfill\rule{0pt}{0pt}
\caption{\label{fig555}Examples of minimalistic linear Vernier patterns, for $k=3$ (left) and for $k=4$ (right).}
\end{figure}

Such Vernier patterns are called \emph{minimal Vernier patterns}. For small $N$ one can even find \emph{minimalistic} Vernier patterns, i.\,e. the patterns with $k(k-1)/2 = \lfloor N/2 \rfloor$ (resp. $k(k-1)/2 = N_1$ for linear patterns).
Figures~\ref{fig2}--\ref{fig555} give some examples of minimalistic circular and linear Vernier patterns. Unfortunately, to the best of our knowledge there are no minimalistic Vernier patterns for $N>12$.

So the problem of finding the \emph{minimal} Vernier patterns may be considered as a serious combinatorial problem. The simplest Vernier pattern described in Section~\ref{sec-vernier-all} yields the obvious upper bound for the parameter $k$: $k \leqslant 2 \sqrt{N}$. One may expect that minimal Vernier patterns should have $k$ much closer to the lower bounds given above.

\section*{Acknowledgments}
This study was supported by a research grant No.~14.Y26.31.0004 from the Government of the Russian Federation (M.\,G.\,Sadovsky) and the grant from Russian Ministry of Education and Science to Siberian Federal University, contract $N^{o}$ 1.1462.2014/K (S.\,P.\,Tsarev). The authors thank Prof. S.\,V.~Znamenskij for useful discussions; the idea of Vernier gauge for acceleration of search was also independently found by him.

\end{document}